# The Spectral and Magnetic Properties of $Er^{3+}$ and $Yb^{3+}$ Ions in $Y_2Ti_2O_7$ Crystals with a Pyrochlore Structure

R. G. Batulin, M. A. Cherosov, I. F. Gil'mutdinov, B. F. Khaliulin, A. G. Kiiamov,
V. V. Klekovkina, B. Z. Malkin, I. R. Mukhamedshin, I. E. Mumdzhi,
S. I. Nikitin, A. A. Rodionov, and R. V. Yusupov

Kazan Federal University, Kazan, 420008 Russia

e-mail: *Roman.Yusupov@kpfu.ru*

**Abstract**

Cubic $Y_2Ti_2O_7$ single crystals doped with $Er^{3+}$ or $Yb^{3+}$ ions have been studied by the methods of electron spin resonance (ESR) and selective laser spectroscopy. ESR spectra exhibit signals from rare-earth ions that substitute for yttrium ions in sites with local trigonal symmetry. The *g* tensor components are determined. The results of optical investigations indicate that impurity centers of several types are formed; the sublevel energies of the ground and excited multiplets of these centers are found. Among the great variety of detected optical centers, the centers that dominate in the formation of ESR spectra are discriminated. An analysis of the experimental data using the exchange-charge model have made it possible to determine the sets of parameters of the crystal field for $Er^{3+}$ and $Yb^{3+}$ ions substituting $Y^{3+}$ ions in regular crystallographic sites in pyrochlore $Y_2Ti_2O_7$.

## 1. INTRODUCTION

In the last two decades, rare-earth (RE) metal compounds with a pyrochlore $R_2M_2O_7$ structure (R is a RE metal and M is $Ge^{4+}$, $Ti^{4+}$, $Pt^{4+}$, or $Sn^{4+}$) containing networks of tetrahedra formed by RE ions with shared vertices have been of great interest for researchers, because of the effects of geometric magnetic frustration observed therein [1]. Pyrochlores containing ions with easy-plane magnetic anisotropy (XY-type ions), which include $Yb^{3+}$ and $Er^{3+}$, take a special place in this series. At low temperatures, a wide spectrum of magnetically ordered states was observed in pyrochlores $Yb_2M_2O_7$ and $Er_2M_2O_7$ (in particular, phases with noncoplanar and coplanar antiferromagnetic order, antiferromagnetic Palmer–Chalker phase, and phase with separated ferromagnetic ordering [2–4]). Specifics of magnetic properties of different RE pyrochlores is determined by the magnetic characteristics of single RE ions and their energy spectrum in the local crystal field.

The structure of energy spectra and *g* factors of the ground state of $Er^{3+}$ and $Yb^{3+}$ ions both in magnetically concentrated and in diluted compounds with a pyrochlore structure were investigated previously by optical [5], neutron [6–8], and Mossbauer [9] spectroscopies. However, the sets of parameters of the crystal field given in the literature differ significantly for both $Er^{3+}$ and $Yb^{3+}$ ions, and the magnetic or spectral characteristics calculated using these parameters are not consistent with measurement data. The purpose of this study was to eliminate the existing contradictions based on experimental and theoretical investigations of the physical properties of isolated impurity RE ions in diamagnetic pyrochlore $Y_2Ti_2O_7$. We synthesized $Y_2Ti_2O_7$ single crystals doped with $Er^{3+}$ or $Yb^{3+}$ ions (0.5 at %) and measured their ESR spectra and selectively excited emission spectra. The data obtained were interpreted basing on the crystal-field simulation. The presented basic sets of parameters of the crystal fields in $Y_2Ti_2O_7{:}Er^{3+}$ and $Y_2Ti_2O_7{:}Yb^{3+}$ crystals can be used to determine physically justified exchange-interactions parameters in magnetoconcentrated systems.

## 2. EXPERIMENTAL

The $Y_2Ti_2O_7$ compounds were synthesized using oxides $Y_2O_3$ and $TiO_2$ with a purity of 99.95% (Alfa Aesar). The powders were taken in a close-to-stoichiometry ratio with a small excess (~0.5 mol %) of titanium oxide. The crystals were doped with RE ions using $Er_2O_3$ or $Yb_2O_3$ salts with impurity contents below 0.05 at %. The mixture was stirred and homogenized by grinding in an agate mortar for 6 h. Then the solid-phase synthesis in air at a temperature of 1000°C was performed for 12 h. The product obtained was checked for the presence of the desired phase with a pyrochlore structure and the absence of foreign phases by X-ray powder diffraction analysis; if necessary, the procedure of solid-phase synthesis was repeated. Further, a cylindrical ingot 8 mm in diameter and 80–100 mm long was formed using a hydraulic press at the hydrostatic pressure of 200 bar. The crystals were grown by the floating-zone method with optical heating on an FZ-T-4000-H-VII-VPO-PC system (Crystal Systems Corp., Japan) with a rate of 2 mm/h. As a result, we obtained $Y_2Ti_2O_7$ crystals containing $Er^{3+}$ and $Yb^{3+}$ ion impurities with a concentration of ~0.5 at %.

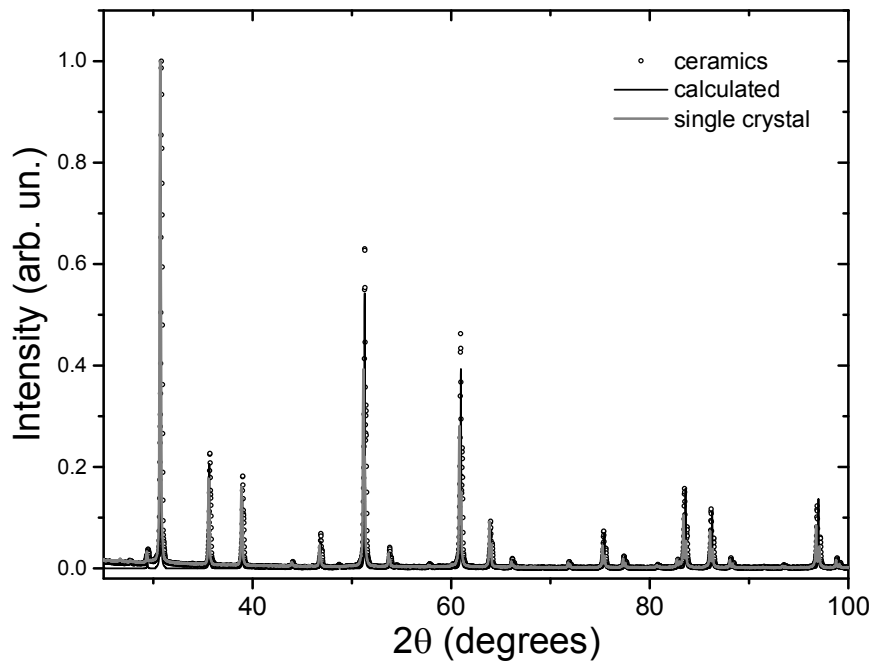

**Fig. 1.** Calculated (lower trace) and measured powder diffraction patterns of the $Y_2Ti_2O_7{:}Er^{3+}$ compound (0.5 at %) after the solid-phase synthesis (ceramics, middle trace) and after the crystal growth (upper trace).

In both cases, the samples had a brown–yellow color, which is not related to their doping with RE ions. The sample coloration is most likely due to the compound composition deviation from stoichiometry and, as a consequence, formation of color centers. To verify this suggestion, we analyzed powder diffraction patterns of the sample after the solid-phase synthesis and a finely-ground piece of the crystal. The measurements were carried out on a Bruker D8 Advance X-ray diffractometer with a source having a copper anticathode. The radiation had a doublet spectrum characteristic of Cu$K_\alpha$ radiation, which causes the corresponding structure of maxima in the measured signals. Figure 1 shows the diffraction pattern calculated using the Powder Cell program [10] and the powder measurement results. The diffraction pattern of the ingot after the solid-phase synthesis reproduces in detail the calculation results. However, the intensity ratio of diffraction maxima in the signal from the powder prepared from the grown crystal changed significantly. Notably, all the expected peaks are present in the diffraction pattern and observed at the expected angles.

The intensities of diffraction maxima change with a change in the crystallographic-site occupancies (in particular, at the vacancy formation in oxygen sublattices). Yttrium oxide has a much higher melting temperature compared to titanium oxide, and the maintaining of the molten state of the crystallization zone for a long time may lead to a change in the compound composition and deviation from stoichiometry. According to the results of electron-dispersion analysis, there is a deficiency in titanium ions in the grown crystals. The atomic concentration ratio of the cations was Y : Ti = 1 : 0.90.

The ESR spectra were measured on a Bruker ESP300 X-range cw spectrometer (~9.5 GHz) equipped with an ER4102ST standard rectangular cavity with the operating $TE_{102}$ mode. The measurements were performed at a temperature of 20 K using an ESR9 continuous-flow helium cryostat (Oxford Instruments). The single-crystal samples were oriented along the crystallographic axes using the X-ray diffractometer equipped with a Euler's craddle. The orientation accuracy was ±2°. The crystallographic directions of the sample were oriented with respect to the dc magnetic field using a sample holder with a capability of its rotation around two perpendicular directions, which allows one to accurately set any direction in the sample.

Selective laser spectroscopy measurements were carried out using the following narrow-band tunable lasers: an LQ109 Ti:Sapphire laser with a laser line-width of 0.03 nm or an LQ125 dye laser with a spectral width of 0.05 nm (Solar LTD, Minsk, Belarus). In both cases, the pumping was performed by the second-harmonic radiation of a neodymium-doped garnet laser with a wavelength of 532 nm. The excitation-pulse width was 10 ns. Cooled FEU-109 (visible and UV ranges) and FEU-83 (near-IR range) photomultiplier tubes operating in the photon counting mode were used for detection. The detected luminescence wavelength was selected by an MDR-23 monochromator with a diffraction grating of 600 lines per millimeter and a spectrometer slit width of 2.6 nm/mm in combination with absorption light filters. The detection system has a possibility of choosing and limiting the time interval of pulse counting after the excitation. The luminescence kinetics was measured by an MCS-Turbo multichannel counter. In these experiments, the samples were placed in a bath helium cryostat and kept at a temperature of 4.2 K.

### 3. MEASUREMENT RESULTS

$Y^{3+}$ ions in a cubic pyrochlore lattice (sp. gr. Fd3m) are in the 16d sites with local trigonal symmetry $D_{3d}$ (3m). Four $Y^{3+}$ ion sublattices form a network of regular tetrahedra with shared vertices (Fig. 2). Note that bond lengths $R_k$ between $Y^{3+}$ ions and oxygen ions differ significantly for the first (two $O^{2-}$ ions on the threefold symmetry axis, $R_1$ = 2.185 A) and second (six $O^{2-}$ ions forming a corrugated hexagon, $R_2$ = 2.51 A) coordination shells.

$Er^{3+}$ and $Yb^{3+}$ impurity ions mainly substitute $Y^{3+}$ ions, because of identical oxidation states and proximity of ionic radii. Four magnetically inequivalent RE-ion centers arise in a cubic crystal. In our single-crystal samples with significant deviations of the chemical composition from stoichiometry, low-symmetry RE-ion centers substituting $Y^{3+}$ ions at defect positions can be formed.

In a trigonal-symmetry crystal field, the ground multiplets $^6I_{15/2}$ of the $Er^{3+}$ ion (electronic configuration $4f^{11}$) and $^2F_{7/2}$ of the $Yb^{3+}$ ion (electronic configuration $4f^{13}$) are split into four and eight doublets, respectively, the wave functions of which are transformed according to the irreducible representations $\Gamma_4$ or $\Gamma_{56}$ of the point symmetry group $D_{3d}$. The magnetic properties of these states differ significantly. The $g$ tensor of the so-called dipole–octupole doublets $\Gamma_{56}$ has only one nonzero component $g_\parallel$ (doublets $\Gamma_{56}$ are split only by the magnetic field directed along the trigonal symmetry axis), whereas both (longitudinal and transverse) components of the $g$ tensor of doublets $\Gamma_4$ are nonzero.

The first excited sublevels of the ground multiplets have an energy of ~50 cm$^{-1}$ for the $Er^{3+}$ ion [8] and ~600 cm$^{-1}$ for $Yb^{3+}$ [5]. At 20 K, only the ground doublets $\Gamma_4$ are predominantly occupied, and the ESR spectra exhibit only transitions between their Zeeman sublevels. Figure 3 shows the ESR spectra of the (a) $Y_2Ti_2O_7$:$Er^{3+}$ and (b) $Y_2Ti_2O_7$:$Yb^{3+}$ crystals with the RE-ion concentration of 0.5 at % measured in magnetic fields $B_0$ oriented along the [111], [110], and [001] crystallographic directions. It can be seen that there are two strong lines in the spectra of both ions in each of the first two cases and one strong line in the third case. This fact suggests that signals from the trigonal-symmetry centers dominate in the ESR spectra.

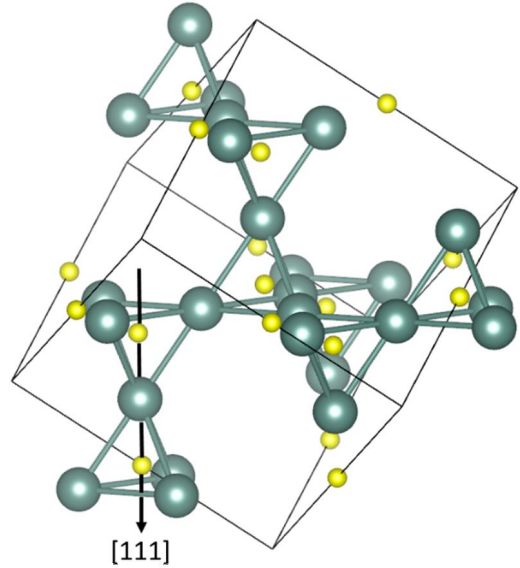

[111]

**Fig. 2.** Network of $Y^{3+}$ ion tetrahedra with shared vertices in the $Y_2Ti_2O_7$ compound with a pyrochlore structure (large circles). A pair of tetrahedra is shown in the lower left corner, which determines the local structure in the $Y^{3+}$ ion site; small circles indicate pairs of nearest oxygen ions.

The effective $g$ factor of the axial center in a magnetic field making angle $\theta$ with the symmetry axis is

$$g = \sqrt{g_\parallel^2 \cos^2\theta + g_\perp^2 \sin^2\theta}. \qquad (1)$$

Two signals are observed in magnetic field $\mathbf{B}_0 \parallel [111]$: one is from a center with the axis directed along the applied field ($g = g_\parallel$) and the other is from three other equivalent centers ($\theta = 70.53°$). Similarly, ESR signals of two pairs of centers with angles $\theta = 90°$ and $35.26°$ are observed in field $\mathbf{B}_0 \parallel [110]$. In field $\mathbf{B}_0 \parallel [001]$, all four centers are equivalent ($\theta = 54.74°$). An analysis of the spectra shown in Fig. 3 made it possible to determine the $g$-tensor components: $g_\parallel = 2.29$ (2.315) and $g_\perp =$

6.76 (6.763) for the $Er^{3+}$ ion; $g_\parallel$ = 1.787 (1.82) and $g_\perp$ = 4.216 (4.21) for the $Yb^{3+}$ ion (*g* factors calculated within the below-presented models of a crystal field are given in parentheses). Note that strong signals in the ESR spectra are due to the transitions between magnetic sublevels of even $Er^{3+}$ and $Yb^{3+}$ ion isotopes (nuclear spin *I* = 0). The spectral lines in the vicinity of the strong components are due to hyperfine interaction for the $^{167}Er$ isotopes (*I* = 7/2, natural abundance 22.09%) or the $^{171}Yb$ (*I* = 1/2, 14.3%) and $^{173}Yb$ (*I* = 5/2, 16.2%) isotopes.

The Stark structure of the multiplets of $Er^{3+}$ and $Yb^{3+}$ impurity ions in the $Y_2Ti_2O_7$ crystals was analyzed by site-selective laser spectroscopy. The method and objects of study have a specificity that hinders significantly the interpretation of experimental data. First, any synthetic crystal is imperfect: its structure, regardless of the synthesis method, always contains defects, both point (cation and anion vacancies, impurities) and extended (dislocations), therefore some impurity ions are found in the sites distorted by the nearby defects. Second, site-selective laser spectroscopy is a high-sensitivity method and allows one to select and investigate luminescent centers, the concentrations of which in samples differ by several orders of magnitude. Therefore, formation of multiple impurity centers in activated crystals hinders identification of observed spectra.

The specificity of $Y_2Ti_2O_7$:$Er^{3+}$ and $Y_2Ti_2O_7$:$Yb^{3+}$ crystals is related to the presence of an inversion center in the $Y^{3+}$ ion site (substituted by RE ions) in the regular lattice. In this case, the *f*–*f* transitions are parity-forbidden within the electric dipole approximation.

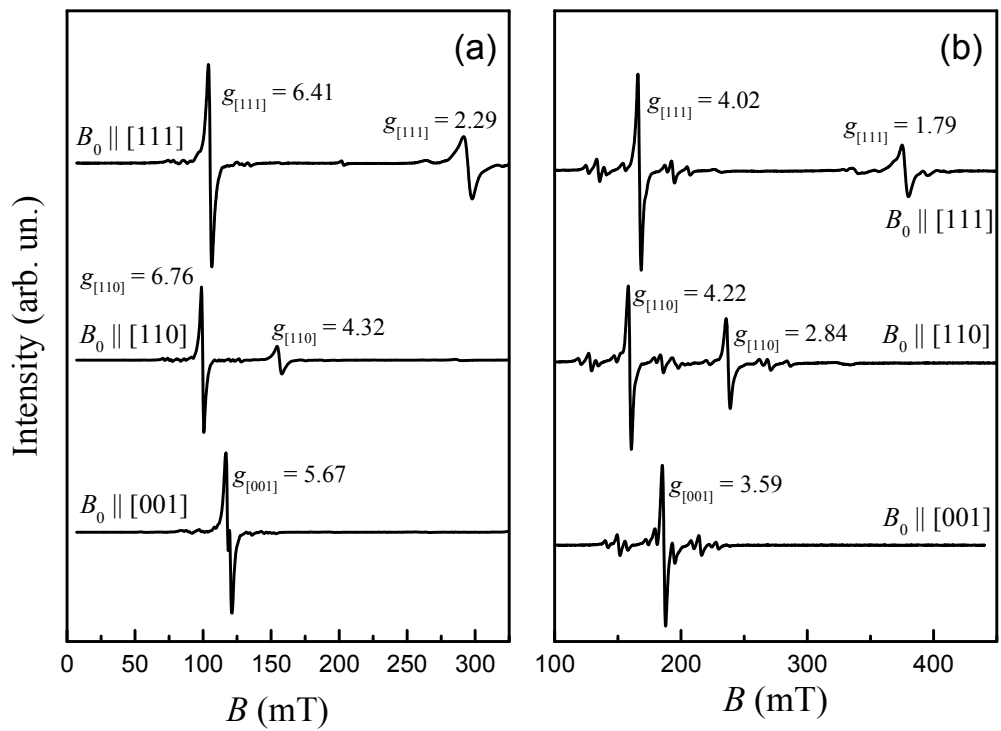

**Fig. 3.** ESR spectra of the (a) $Y_2Ti_2O_7$:$Er^{3+}$ and (b) $Y_2Ti_2O_7$:$Yb^{3+}$ crystals measured at the given directions of magnetic field $B_0$ at *T* = 20 K and frequency ν = 9.356 GHz.

Transitions from the ground multiplet $^4I_{15/2}$ of $Er^{3+}$ ions to sublevels of most of the multiplets in the visible and near-IR ranges are also forbidden within the magnetic dipole approximation (|Δ*J*| > 1). Since the parityforbiddenness can be removed by a nearby defect, the probabilities of optical transitions of $Er^{3+}$ ions in regular and defect-associated sites may differ by several orders of magnitude. In this case, the intensity of the observed luminescence and excitation spectra is not an objective measure of the center concentration; the differentiation of regular and defect centers may require additional information. The specificity of impurity centers formed by $Yb^{3+}$ ions is somewhat different; it is related to traditionally intense vibronic structure of optical spectra of these ions. In this case, separation of the lines corresponding to zero-phonon and vibronic transitions is a complex problem.

Note that, in low-temperature experiments, the information about energy-sublevel diagram of the ground and excited multiplets is provided by the luminescence and excitation spectra, respectively.

Let us now consider the obtained experimental results. Figure 4 shows the strongest luminescence spectra from the lower sublevel of the multiplet $^4S_{3/2}$ to the ground multiplet $^4I_{15/2}$ of three different impurity centers of $Er^{3+}$ ions. The line strengths of the A and B centers are almost identical, while the C-center line strength is several times lower. The lifetimes for three center types are approximately equal (on the order of several hundred microseconds). The problem of determining the center corresponding to $Er^{3+}$ ions in regular sites of $Y^{3+}$ ions was solved by comparing the structures of sublevels of the ground multiplet (that manifest themselves in the luminescence spectra) and the structure found in experiments on inelastic neutron scattering [8] (see Table 1).

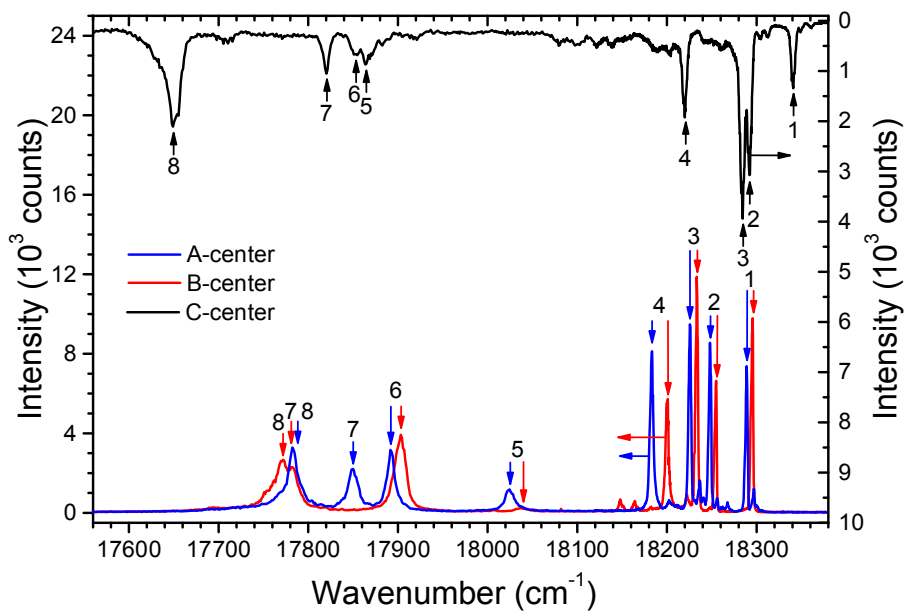

**Fig. 4.** Luminescence spectra of the three types of the impurity $Er^{3+}$ ion centers in the $Y_2Ti_2O_7$ crystal possessing the strongest intensity; wave numbers of the excitation light are $v_A = 20\ 430.7$ cm$^{-1}$, $v_B = 19\ 004.2$ cm$^{-1}$, and $v_C = 20\ 677.4$ cm$^{-1}$; $T = 4.2$ K.

Figure 5 shows the luminescence and excitation spectra of the ground (intensity-dominant in the integral spectra) impurity center of $Yb^{3+}$ ions in the $Y_2Ti_2O_7$ crystals. Along with the regular center, four more types of centers were distinguished, which are characterized by the line strength that is several orders of magnitude lower. The luminescence lifetime was 3.59 ms for the regular center and varied from 2.06 to 3.54 ms for the other centers. A characteristic feature of the luminescence and excitation spectra of the regular $Yb^{3+}$ center is a rich vibrational structure. The spectra shown in Fig. 5 exhibit high symmetry. In this case, to identify zero-phonon lines in the excitation spectrum, we compared it with the absorption spectrum of the concentrated $Yb_2Ti_2O_7$ compound [5]. The line positions in the luminescence spectrum were also determined by comparing with the data of experiments on inelastic neutron scattering in $Yb_2Ti_2O_7$ [6].

## 4. RESULTS AND DISCUSSION

Simulation of the measured optical and ESR spectra of $Er^{3+}$ ions was performed using the parameterized single-ion Hamiltonian

$$H = H_{FI} + H_{CF} + H_Z, \qquad (2)$$

determined in the complete basis of 364 states of the $4f^{11}$ electron shell. Here, $H_{FI}$ is the standard Hamiltonian of a free ion [11], $H_{CF}$ is the energy of $4f$ electrons in a trigonal-symmetry crystal field determined by six independent parameters $B_p^n$ (the local coordinate system with the $z$ axis directed along the symmetry axis of the center under consideration is used):

$$H_{CF} = B_2^0 O_2^0 + B_4^0 O_4^0 + B_4^3 O_4^3 + B_6^0 O_6^0 + B_6^3 O_6^3 + B_6^6 O_6^6, \qquad (3)$$

$O_p^n$ are the linear combinations of spherical tensor operators [12], and $H_Z = -\mu B$ is the electron energy in magnetic field $\mathbf{B}$ ($\mu = -\mu_B(k\mathbf{L} + 2\mathbf{S})$ is the magnetic moment; $\mathbf{L}$ and $\mathbf{S}$ are, respectively, the orbital and spin moments of the RE ion; $k$ is the orbital-reduction factor; and $\mu_B$ is the Bohr magneton).

**Table 1.** Measured and calculated energies of different states (in $cm^{-1}$) of RE ions in the $Y_2Ti_2O_7{:}Er^{3+}$, $Y_2Ti_2O_7{:}Yb^{3+}$, $Er_2Ti_2O_7$, and $Yb_2Ti_2O_7$ crystals.

| | $Er^{3+}$ ($^4I_{15/2}$) | | | | | $Yb^{3+}$ | | |
|---|---|---|---|---|---|---|---|---|
| Γ | ErTiO experimental [8] | YTiO:Er experimental [This study] | YTiO:Er calculation [This study] | $^{2S+1}L_J$ | Γ | YbTiO experimental [6] | YTiO:Yb experimental [This study] | YTiO:Yb calculation [This study] | |
| $Γ_4$ | 0 | 0 | 0 | $^2F_{7/2}$ | $Γ_4$ | 0 | 0 | 0 | |
| $Γ_{56}$ | 50.8 | 49.3 | 47.9 | | $Γ_4$ | 617.9 | - | 613 | 541* |
| $Γ_4$ | 58.8 | 56.8 | 57.9 | | $Γ_{56}$ | 658.6 | 655 | 657 | 590* |
| $Γ_4$ | 126.5 | 121.8 | 132.4 | | $Γ_4$ | 936.3 | 976 | 966 | 833* |
| $Γ_4$ | 484.9 | 478.8 | 461.9 | $^2F_{5/2}$ | $Γ_4$ | - | 10291 | 10291 | 10335* |
| $Γ_{56}$ | 501.9 | 490.3 | 473.3 | | $Γ_4$ | - | 10856 | 10852 | 10778* |
| $Γ_4$ | 534.1 | 522.4 | 504.0 | | $Γ_{56}$ | - | 11025 | 11042 | 10968* |
| $Γ_{56}$ | 702.4 | 694.7 | 700.3 | | | | | | |

* Indicates that the hybridization with charge-transfer states is disregarded (see text).

The initial parameters of the crystal field were calculated within the exchange-charge model [12, 13] and then corrected by comparing the calculated and measured spectral characteristics of the $Y_2Ti_2O_7{:}Er^{3+}$ crystal. The obtained set of parameters is listed in Table 2; the energies of multiplet sublevels (Table 1) and $g$ factors of $Er^{3+}$ ions at regular-lattice sites, calculated using these parameters, are in good agreement with the experimental data.

However, when considering the spectral characteristics of the $Y_2Ti_2O_7{:}Yb^{3+}$ crystal, we found that the Stark structure of the multiplets $^2F_J$ ($J$ = 7/2 or 5/2) of $Yb^{3+}$ ions cannot be reproduced within the standard crystal-field approximation using physically justified parameters that are comparable with the parameters of crystal fields in other RE pyrochlores. According to the previously developed model of the electronic structure of $Yb^{3+}$ ions in complexes with low-energy charge-transfer band [14], we assume that a significant contribution to the formation of the observed spectrum is from hybridization of the ground electronic configuration $4f^{13}$ with the excited configurations $4f^{14}–2p^5$ formed at the virtual charge transfer from the $2p^6$ shells of the two nearest ligands to the unfilled shell of the $Yb^{3+}$ ion.

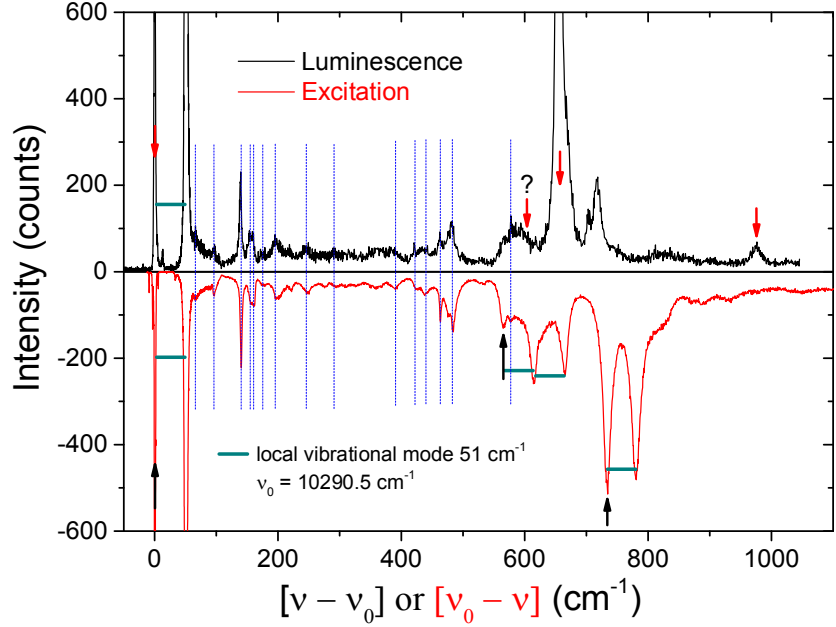

**Fig. 5.** Luminescence (upper panel) and excitation (lower panel) spectra of the $Y_2Ti_2O_7$:$Yb^{3+}$ crystal that are energy shifted for comparison. The arrows indicate lines assigned to the zero-phonon transitions and the dotted lines are vibrational satellites entering both spectra.

The Hamiltonian of the $Yb^{3+}O_2^{2-}$ complex operating in the space of 14 $Yb^{3+}$ ion states ($4f^{13}$) and charge-transfer states (six spin orbitals of each of the two nearest $O^{2-}$ ions with a hole on the $2p$ shell) can be written as

$$H = H(4f^{13}) + H(ch-t) + H(4f^{13}, ch-t). \qquad (4)$$

The operator $H(4f^{13})$ includes the spin–orbit interaction energy and interaction energy between a hole on the $4f$ shell and the crystal field (see (3)). The charge-transfer state energies are determined by the Hamiltonian $H(ch-t)$ containing energy difference $\Delta$ between the $2p$ hole on the ligand and the hole on the $4f$ shell, spin–orbit interaction energy of the $2p$ hole, and its interaction with the crystal field (since the ligands under consideration are at the sites with tetrahedral symmetry in the regular lattice, the crystal field at the $2p$ hole is formed only by the electron transferred from the ligand to the RE ion). The operator $H(4f^{13}, ch-t)$ relates the charge-transfer states with the wave functions of the $Yb^{3+}$ ion. The matrix of this operator was parameterized by two hopping integrals $T_\sigma$ and $T_\pi$ for the $\sigma$ and $\pi$ bonds, respectively, on the assumption that the degree of hybridization of the $4f$ orbitals of the RE ion and $2p$ orbitals of the ligands is proportional to their overlap integrals.

**Table 2.** Parameters of the crystal fields (in cm$^{-1}$) exerted on RE ions in the $Y_2Ti_2O_7$:$Er^{3+}$, $Y_2Ti_2O_7$:$Yb^{3+}$, $Er_2Ti_2O_7$, and $Yb_2Ti_2O_7$ crystals.

| | | $Er^{3+}$ | | | $Yb^{3+}$ | |
|---|---|---|---|---|---|---|
| $p\ k$ | $Er_2Ti_2O_7$ [5] | $Er_2Ti_2O_7$ [8] | $Y_2Ti_2O_7$:Er This study | $Yb_2Ti_2O_7$ [6] | $Y_2Ti_2O_7$:Yb [5] | $Y_2Ti_2O_7$:Yb This study |
| 2  0 | 267.0 | 210.9 | 239.8 | 278.9 | 273 | 264.8 |
| 4  0 | 315.5 | 328.6 | 311.8 | 274.8 | 317.5 | 270.8 |
| 4  3 | -2212.6 | -2766.8 | -2305.2 | -1514.6 | -1780.7 | -2155.2 |
| 6  0 | 43.6 | 45.3 | 45.7 | 59.3 | 52.5 | 44.9 |
| 6  3 | 667.3 | 621.9 | 666.6 | 2213.7 | 591.8 | 636.6 |
| 6  6 | 717.2 | 783.3 | 753.2 | 395.4 | 719.1 | 683.2 |

The values of three variable parameters of the model ($T_\sigma$ = 1175 cm$^{-1}$, $T_\pi$ = –600 cm$^{-1}$, and $\Delta$ = 24200 cm$^{-1}$) were obtained by comparing the calculated energies of the $Yb^{3+}O_2^{2-}$ complex and the $g$-tensor components of its ground state with the measurement results. The crystal-field parameters in operator (3) that were used in the calculations are given in Table 2; they do not contradict the sets of parameters obtained from the analysis of spectra of other RE pyrochlores (in particular, they are in agreement with our parameters of the crystal field in the $Y_2Ti_2O_7$:$Er^{3+}$ crystal).

## 5. CONCLUSIONS

Based on the performed measurements of ESR spectra and selectively excited luminescence spectra, the values of $g$ factors were determined and the energy-level diagrams of single $Er^{3+}$ and $Yb^{3+}$ ions in the $Y_2Ti_2O_7$ crystal with a pyrochlore structure were refined. An analysis of the experimental data performed within the exchange-charge model yielded the self-consistent description of the magnetic and spectral properties of $Er^{3+}$ and $Yb^{3+}$ ions substituting $Y^{3+}$ ions in regular crystallographic sites and made it possible to determine the corresponding sets of parameters of the crystal fields.

To describe the energy spectrum of $Yb^{3+}$ ions, a new interpretation of the crystal field was proposed based on a quantitative hybridization model for a hole in the $4f$ ytterbium shell and $2p$ holes at two anomalously close oxygen ions. This model may be useful for the explanation of the high sensitivity of $Yb^{3+}$ ions to lattice strain due to the nonstoichiometry of pyrochlores. The possibility of refining the parameters of anisotropic exchange interactions between RE ions in the concentrated $Yb_2Ti_2O_7$ and $Er_2Ti_2O_7$ systems was discovered.

V.V. Klekovkina and B.Z. Malkin acknowledge the support of the Russian Foundation for Basic Research, project no. 17-02-00403; I.E. Mumdzhi and R.V. Yusupov acknowledge the support by the subsidy allocated to Kazan Federal University (KFU) to fulfil the state assignment in the sphere of scientific activities, tasks nos. 3.6722.2017/8.9 and 3.7704.2017/4.6.